%% file: main.tex
\documentclass[prl,aps,amsmath,twocolumn,amssymb,nofootinbib,preprintnumbers,superscriptaddress]{revtex4-2}
\usepackage{graphicx,array}
\usepackage{hyperref}
\usepackage{xcolor}
\usepackage{amsmath,amssymb,slashed,latexsym}
\usepackage{bm,xspace}
\usepackage{braket}
\usepackage{placeins}
\usepackage{adjustbox}
\usepackage[caption=false]{subfig}
\usepackage{enumitem}
\usepackage{amsthm}
\usepackage{tikz}
\usetikzlibrary{positioning, arrows.meta,calc}

\def\eb{\ensuremath{E_b}\xspace}
\def\pcz{\ensuremath{\mathcal{Z}}\xspace}
\def\pczv{\ensuremath{\vec{\mathcal{Z}}}\xspace}
\def\pcn{\ensuremath{\mathcal{N}}\xspace}
\def\pcnv{\ensuremath{\vec{\mathcal{N}}}\xspace}

\definecolor{darkblue}{cmyk}{1,0.4,0,0.3}
\definecolor{violet}{cmyk}{0,1,0,0.2}
\hypersetup{colorlinks, bookmarksnumbered, citecolor=darkblue, linkcolor=darkblue, pdfstartview=FitH, urlcolor=darkblue, linktocpage}

\begin{document}

\preprint{CERN-TH-2025-153}
\title{The DNA of nuclear models: How AI predicts nuclear masses}

\author{Kate A. Richardson}\email{karich@mit.edu}
\affiliation{NSF AI Institute for Artificial Intelligence and Fundamental Interactions} \affiliation{Laboratory for Nuclear Science, MIT, Cambridge, MA 02139, USA}
\author{Sokratis Trifinopoulos}\email{{sokratis.trifinopoulos@cern.ch}}
\affiliation{NSF AI Institute for Artificial Intelligence and Fundamental Interactions} \affiliation{Laboratory for Nuclear Science, MIT, Cambridge, MA 02139, USA}
\affiliation{Theoretical Physics Department, CERN, Geneva, Switzerland}
\affiliation{Physik-Institut, Universit\"at Z\"urich, 8057 Z\"urich, Switzerland}
\author{Mike Williams}\email{mwilliams@mit.edu}
\affiliation{NSF AI Institute for Artificial Intelligence and Fundamental Interactions} \affiliation{Laboratory for Nuclear Science, MIT, Cambridge, MA 02139, USA}

\begin{abstract}
Obtaining high-precision predictions of nuclear masses, or equivalently nuclear binding energies, $E_b$, remains an important goal in nuclear-physics research.
Recently, many AI-based tools have shown promising results on this task, some achieving precision that surpasses the best physics models.
However, the utility of these AI models remains in question given that predictions are only useful where measurements do not exist, which inherently requires extrapolation away from the training (and testing) samples. 
Since AI models are largely \textit{black boxes}, the reliability of such an extrapolation is difficult to assess. 
We present an AI model that not only achieves cutting-edge precision for $E_b$, but does so in an interpretable manner. 
For example, we find that (and explain why) the most important dimensions of its internal representation form a double helix, where the analog of the hydrogen bonds in DNA here link the number of protons and neutrons found in the most stable nucleus of each isotopic chain. 
Furthermore, we show that the AI prediction of $E_b$ can be factorized and ordered hierarchically, with the most important terms corresponding to well-known symbolic models (such as the famous liquid drop). 
Remarkably, the improvement of the AI model over symbolic ones can almost entirely be attributed to an observation made by Jaffe in 1969 based on the structure of most known nuclear ground states.
The end result is a fully interpretable data-driven model of nuclear masses based on physics deduced by AI. 
\end{abstract}

\maketitle

Atomic nuclei consist of $Z$ protons and $N$ neutrons bound together by the strong nuclear force. 
Even though the nucleus was discovered over a century ago---and quantum chromodynamics 
more than 50 years ago---first principles calculations of nuclear masses, or equivalently binding energies, \eb, are still only possible for the smallest nuclei. 
Notably, many open problems in nuclear and (astro)particle physics are limited by a lack of precise knowledge of nuclear masses, either directly or indirectly via other quantities which require them as inputs.\footnote{Examples of physical phenomena that necessitate precise predictions of nuclear masses include $r$-process nucleosynthesis~\cite{Burbidge:1957vc}, the nuclear neutron skin and its implications on the structure of neutron stars~\cite{Brown:2000pd, Horowitz:2000xj, Gandolfi:2011xu}, the exploration of the boundaries of the nuclear landscape~\cite{Erler2012TheLO}, and  exotic phenomena such as halo nuclei~\cite{Nortershauser:2008vp} and shape coexistence~\cite{Heyde:1983zz,Wood:1992zza}.}
Experimentally, precise measurements have been made for the masses of (quasi)stable nuclei~\cite{Wang:2021xhn}; however, measurements of highly unstable nuclei are currently challenging, and thus,   
%
must be predicted using some combination of tractable theoretical calculations, {\em e.g.}\ using phenomenological potentials, and empirical observations of other nuclei. 
Despite achieving an impressive level of precision, even the best such model is not sufficient to solve many open problems, {\em e.g.}, $r$-process nucleosynthesis~\cite{Arnould:2007gh,Martin:2015xql, Mumpower:2015ova}. 

%

The need for higher precision has motivated the use of artificial intelligence (AI) to predict nuclear masses~\cite{Zhang:2017zvb,Niu:2018csp,Wu:2021hil,Niu:2022gwo,Yuksel:2021nae,Zhao:2022hqr,Wu:2022nnc,Zeng:2022azv,Bobyk:2022xpc,Lovell:2022pkw,Kitouni:2023rct,Kitouni:2024ulw,Choi:2024ypo,Lu:2024nkr,Sharma:2024rdl,Jalili:2025nsh,Bentley:2025fbj,Ye:2025fkr}.
These studies have varied in approach from taking the best physics models as inputs and only learning to predict tiny corrections to being fully data-driven with no {\em a priori} physics inputs. 
Excellent performance has been achieved across this spectrum of AI approaches, with the best models improving the precision by about a factor of two over the best physics model for (quasi)stable nuclei (see, {\em e.g.},  Ref.~\cite{Kitouni:2023rct}).
However, the utility of these AI models remains in question given that predictions are only useful where measurements do not exist, {\em i.e.}\ for highly unstable nuclei, which inherently requires extrapolation away from the training (and testing) samples. 
Since most AI models are \textit{black boxes}, the reliability of such an extrapolation is difficult to assess, leaving a vital question unanswered: \textit{What is AI learning?} 

\begin{figure}[t]
    \centering    \includegraphics[width=1\linewidth]{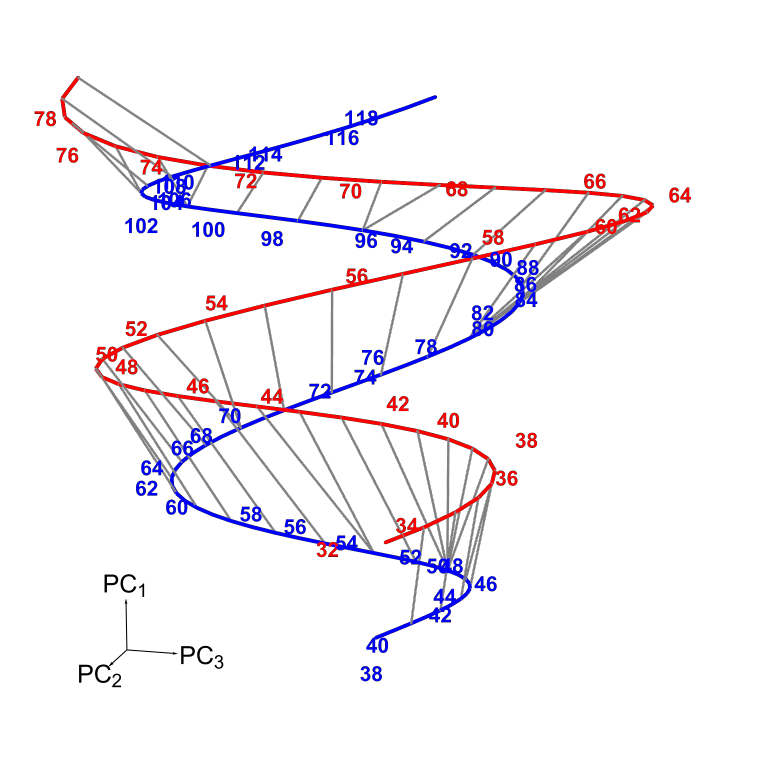}
    \vspace{-4.0em}
    \caption{The three most important \textit{principal components} of the internal representations learned by our AI model for (red) $Z$ and (blue) $N$. The links connect the values of $Z$ and $N$ found in the most stable nucleus of each isotopic chain. Shown is only the range of $(Z,N)$ values where \eb is well described by the volume and asymmetry terms in Eq.~\eqref{eq:SEMF}, with only even values labeled to avoid clutter. The curves show fits of helices to the $Z$ and $N$ representations.}
    \label{fig:DNA}
\end{figure}

In this Letter, we present an AI model---a slightly simplified version of the model from Refs.~\cite{Kitouni:2023rct,Kitouni:2024ulw}---that not only achieves cutting-edge precision for predicting nuclear masses, but does so in an interpretable manner. 
For example, we find (and explain why) that the most important dimensions of its internal representation form a double helix (see Fig.~\ref{fig:DNA}), where the analog of the hydrogen bonds in DNA here link the number of protons and neutrons found in the most stable nucleus of each isotopic chain. 
Furthermore, we show that the AI prediction of \eb can be factorized and ordered hierarchically, with the most important terms corresponding to well-known symbolic models, such as the famous liquid drop (LD)~\cite{Weizsacker:1935bkz,Bethe:1936zz}. 
Remarkably, we find that the improvement of the AI model over symbolic ones can almost entirely be attributed to an observation made by Jaffe and collaborators in 1969~\cite{GARVEY:1969xgo}.\footnote{The relevant sections in Ref.~\cite{GARVEY:1969xgo} were the outcome of Jaffe's \textit{Junior Paper} at Princeton (private communication).}
The end result is 
a
model of nuclear masses
based on physics deduced by AI.

Before discussing the AI model, we will quickly review the history and status of physics models that predict \eb. 
The LD model, proposed almost a century ago, treats the nucleus as a highly dense incompressible fluid, thus 
\begin{align}
\label{eq:SEMF}
\eb^{\rm LD} \approx \alpha_v A - & \alpha_s A^{2/3} - \alpha_c\frac{Z(Z-1)}{A^{1/3}} \nonumber  \\
& -\alpha_a \frac{(N-Z)^2}{A} + \alpha_p \frac{\delta(Z,N)}{A^{1/2}}\, ,
\end{align}
where $A = Z+N$ is the total number of nucleons (protons and neutrons) in the nucleus. 
The first three terms account for: 
(volume) the nuclear energy being proportional to the volume of the incompressible fluid due to the short-range nature of the strong force, 
(surface) a correction to the previous term due to nucleons near the surface having fewer nearby nucleons to interact with,
and 
(Coulomb) the electromagnetic potential energy of each pair of protons. 
The last two terms, asymmetry and pairing, both arise due to the Pauli Exclusion Principle. 
The pairing term always favors $Z$ and $N$ being even over odd, though the precise form of $\delta(Z,N)$ varies. 
Despite its simplicity, after fitting the $\alpha_i$ coefficients to data, the LD predictions achieve a typical precision of $\mathcal{O}(\%)$ for moderate nuclei and $\mathcal{O}(0.1\%)$ for large nuclei~\cite{Kirson:2008yvv}, which is, {\em e.g.}, about a factor of 30 from the precision required for $r$-process calculations~\cite{Mumpower:2015ova}. 
Figure~\ref{fig:res-LD-WS4} in our Supplemental Material~\cite{Supp} compares $\eb^{\rm LD}$ to data. 

Over the decades, many additional terms, along with modifications to the terms in Eq.~\eqref{eq:SEMF}, have been proposed, though there is no universally accepted improved symbolic formula.
This is in part due to the near degeneracy of many proposed functions of $Z$ and $N$ over the small region of the $(Z,N)$ plane where measurements exist.
Roughly, the precision can be improved by a factor of two by increasing the number of parameters by a factor of three (relative to the basic LD model)~\cite{Kirson:2008yvv}, which is still an order of magnitude away from the target precision. 

The non-AI models that make the most precise predictions of \eb  employ a \textit{microscopic-macroscopic} approach~\cite{Moller:1993ed,Wang:2010dm, Liu:2011ama,Moller:2012pxr,Wang:2014qqa,Moller:2015fba}.
The microscopic part refers to calculations of single-nucleon energy levels using the approximation that all nucleons experience a mean-field potential due to their interactions with all other nucleons in the nucleus~\cite{Strutinsky:1966bz,Walecka:1974qa,Lalazissis:1996rd,Lalazissis:1999zz,Goriely:2001zz,Bender:2003jk,Geng:2005yu,Chen:2014mza,Chen:2014sca}. 
The potential used is phenomenological in nature with a number of parameters determined by fitting the available nuclear data. 
The macroscopic part refers to how these single-nucleon energy levels collectively vary across the nuclear $(Z,N)$ plane, {\em e.g.}, following an equation similar to Eq.~\eqref{eq:SEMF}. 
Therefore, in this family of models, the macroscopic part is a symbolic expression, while the microscopic part is the output of a simplified many-body quantum-mechanical calculation. 
Currently, version 4 of the Weizs\"acker-Skyrme (WS4) model provides the best precision with a root-mean-square (RMS) error of about 0.28~MeV~\cite{Wang:2014qqa}.\footnote{All RMS values quoted in this Letter are calculated on the same set of nuclei as described in the Supplemental Material.} 
The macroscopic expression, provided in Eq.~\eqref{eq:WS4}, is a slightly modified version of Eq.~\eqref{eq:SEMF} that has 9 parameters.
The potential used in the microscopic calculations is an \textit{axially deformed Woods-Saxon} potential, see Ref.~\cite{Wang:2014qqa} for details. 
Figure~\ref{fig:res-LD-WS4} compares the WS4 predictions to experimental data.

In Ref.~\cite{Kitouni:2023rct}, we (and a few colleagues) presented an AI model that predicts various nuclear observables, including \eb, separation energies, and charge radii.
The model was trained using a multi-task learning approach, where a common internal representation was trained to predict the full set of nuclear observables by leveraging their joint information.
No inductive bias or physics inputs were used, thus these AI predictions are entirely empirical, {\em i.e.}\ they are based only on patterns learned directly from the available nuclear data.
Despite this lack of physics awareness, our AI model was able to achieve an amazing precision of 0.13~MeV, a mean relative precision of $\mathcal{O}(10^{-4})$.
The primary focus of Ref.~\cite{Kitouni:2023rct}, and a follow-up work~\cite{Kitouni:2024ulw}, was AI-centric, specifically how simultaneously learning to perform multiple tasks can improve performance, and that high-dimensional neural networks (NNs) can learn interpretable low-dimensional representations; whereas, our focus here is physics-centric, namely is what the AI model learned useful for science?

To facilitate extracting what knowledge the AI has learned, we study a simplified version of the model in Refs.~\cite{Kitouni:2023rct,Kitouni:2024ulw} that is only trained to predict \eb and only has access to \eb measurements. 
There are some aspects about the model architecture and learning dynamics that are important in understanding what the model has learned.
We present these here in a physics-centric way, with the full details in Ref.~\cite{Supp}. 
The only inputs for each nucleus are $Z$ and $N$, which we promote to 1024-dimensional vectors that are fed into a NN, $F_{\rm{nn}}$, with learnable parameters $\vec{\theta}$:
\begin{align}
Z \to \vec{Z}, N \to \vec{N} \Rightarrow \eb = F_{\rm{nn}}(\vec{Z},\vec{N},\vec{\theta}) \, .
\end{align}
The vectors $\vec{Z},\vec{N}$ are initialized randomly and learn to embed properties about each number of protons and neutrons through training. This includes relative properties which we will show can be stored, {\em e.g.}, in the angles between components of the vectors. 
The objective during training is to minimize the following loss function:
\begin{align}
\label{eq:loss}
\mathcal{L} \!=\! \sum_i \left(E_{b,i}^{\rm ex} \!-\! E_{b,i}^{\rm ai}\right)^2\! +\! \lambda \!\left[ \sum_j Z_j^2 \!+\! \sum_k N_k^2 \!+\! \sum_{\ell} \theta_{\ell}^2  \right] \!, 
\end{align}
where $E_{b,i}^{\rm ex}$ and $E_{b,i}^{\rm ai}$ are the experimental and AI-predicted \eb values for each nucleus in the training data, and $\lambda$ is a hyperparameter that controls the balance between prediction quality and model complexity. 

From a physics perspective, this AI model is a system with degrees of freedom $\{ Z_j, N_k, \theta_{\ell}\}$ initialized in a random configuration, where the training procedure implements a dissipative process to find the ground state, defined as the one with the minimum $\mathcal{L}$. 
The regularization term in Eq.~\ref{eq:loss}, the $\lambda[\ldots]$ term, encourages efficient storage of information in the components of $\vec{Z}, \vec{N}$.
In essence, it attaches to each AI model degree of freedom an oscillator with zero natural length and spring constant $\sqrt{2}\lambda$. 
We can rank directions in $\vec{Z}, \vec{N}$ space by how much projections to those axes affect the \eb predictions using a technique known as principal component analysis (PCA). 
Figure~\ref{fig:DNA} shows the three most important principal components (PCs) of $\vec{Z}, \vec{N}$, denoted $\mathcal{Z}_{1,2,3}, \mathcal{N}_{1,2,3}$, 
%
where a clear double-helix structure has formed. 

There is a resemblance between the structure formed in $\mathcal{Z}$-$\mathcal{N}$ space to that of DNA, 
%
which forms a double helix to minimize energy. 
The nucleotide bases in DNA are hydrophobic, and pushed inwards by the surrounding water in the cell. 
In our AI model, the regularization term provides the force that attracts the elements inwards. 
In DNA, Van der Waals forces oppose the hydrophobic pressure, whereas in our AI model this effect is driven by the need to fit the data well to minimize the error term in $\mathcal{L}$, {\em i.e.}\ to provide good predictions of \eb. 
The volume term is dominant in Eq.~\eqref{eq:SEMF}, hence $\eb \approx \alpha_v A$ and our AI model must be able to build $A$ from $\vec{\mathcal{Z}}$ and $\vec{\mathcal{N}}$.
This drives the most important PCs to be (see Fig.~\ref{fig:DNA} vertical)
\begin{align}
\mathcal{Z}_1 \approx \beta Z, \mathcal{N}_1 \approx \beta N \to A \approx (\mathcal{N}_1 + \mathcal{Z}_1)/\beta \, ,
\end{align}
where $\beta$ is a scale factor and the addition operation is trivial for the downstream NN to implement. 
To produce the volume term, the NN needs to store the factor $\alpha_v/\beta$ in its $\vec{\theta}$ parameters. 
Note that while the regularization term on $\{Z_j, N_k\}$ wants to drive $\beta \to 0$, this results in a contribution from the $\{\theta_{\ell}\}$ regularization of $\alpha_v/\beta \to \infty$. 
As we show in detail in the Supplemental Material, there is a non-zero value of the scale of $\mathcal{Z}_1,\mathcal{N}_1$, that minimizes $\mathcal{L}$ for a given set up, thus compressive regularization pressure is, as expected, counterbalanced by the goodness-of-fit term in $\mathcal{L}$. 

The same reasoning that prevents the first PC from collapsing under the regularization pressure also applies to the second and third PC dimensions shown in Fig.~\ref{fig:DNA}.
However, this does not explain the origins of the helical structure.
Naively, there are no cyclic or oscillatory terms in Eq.~\eqref{eq:SEMF}, making it surprising that the second and third most important PCs form an oscillatory structure. 
The explanation lies in the second most important term in Eq.~\eqref{eq:SEMF}, the asymmetry term, which contains a factor of $(Z-N)^2$.
This term is invariant under $Z \to Z+\gamma, N \to N+\gamma$, {\em i.e.}\ translations along the $\hat{\pcz}_1$ and $\hat{\pcn}_1$ axes (which are nearly aligned) in Fig.~\ref{fig:DNA}. 
When computing  $(Z-N)^2$, only the relative difference of $Z$ and $N$ matters; therefore, the same solution can be repeated all along the first PC axis. 
We show in the Supplemental Material precisely how the double helix is used to obtain $(Z-N)^2$ efficiently.
Electromagnetic interactions of protons break the symmetry between $Z$ and $N$, which is why the double helix in Fig.~\ref{fig:DNA} is not symmetric. 
A symmetric double helix is obtained when the experimental data is replaced with simulated isospin-symmetric data~\cite{Supp}.

\begin{figure*}[t]
    \centering
    \includegraphics[width=\linewidth]{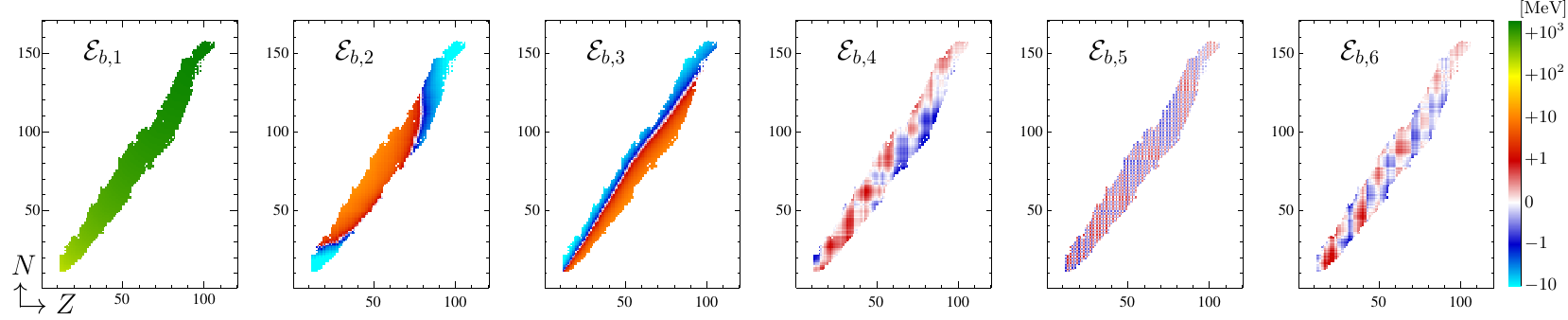}
    \caption{The six most important PCs of the AI \eb predictions. The scale is linear on $[-1,1]$ and logarithmic otherwise. The first three PCs are approximately smooth functions and largely correspond to macroscopic terms, while all lesser PCs are discrete functions, with most approximately factorizing as $F_Z(Z) + F_N(N)$.}
    \label{fig:penultimate-pcs}
\end{figure*}

An advantage of our AI approach---embedding information about proton and neutron number into learnable vectors---is that structures formed in the PC space correlate strongly with predictive power of the model~\cite{Kitouni:2024ulw}.
The PCs are defined such that the most-important components of \pczv and \pcnv by construction affect the most nuclei; structure in these components maps to the macroscopic terms in physics models. 
The fact that there are preferred directions entering the NN, corresponding to the most important \pczv and \pcnv directions, results in a hierarchy forming in the penultimate NN layer, where the terms that will be summed to make the \eb predictions are built. 
Figure~\ref{fig:penultimate-pcs} shows that the most-important PCs in this space, denoted $\mathcal{E}_{b,i}$, are approximately smooth functions; in fact, the first PC is well described by Eq.~\eqref{eq:SEMF}~\cite{Supp}. 
The first three PCs, therefore, largely describe macroscopic terms.
The lesser PCs correspond to microscopic terms, providing corrections for various energy levels.

The form of the lesser PCs in Fig.~\ref{fig:penultimate-pcs} is worthy of discussion.
Recall that the microscopic terms in physics models are calculated using a mean-field many-body quantum-mechanical approach.
However, the AI model has instead, to a good approximation, learned a much simpler estimate for microscopic terms:
\begin{align}
\label{eq:mic-fac}
\eb^{\rm mic} \approx F_Z(Z) + F_N(N) \, ,
\end{align}
where $F_{Z,N}$ are discrete functions of only $Z$ or $N$, and importantly, not both; {\em i.e.}\ the microscopic terms approximately factorize, making them easy to learn. 
%

Remarkably, this same factorization observation was made in 1969~\cite{GARVEY:1969xgo}, though it seems to have been 
forgotten 
until rediscovered by our AI model.
Ref.~\cite{GARVEY:1969xgo} was based on the Garvey-Kelson (GK) relations, patterns found in nuclear data where combinations of \eb from a set of nuclei  approximately cancel almost anywhere in the nuclear plane (examples in \cite{Supp}).
%
%
The GK relations arise because single-nucleon energy levels do not typically vary much in a small region of the nuclear plane. 
%
Any contribution to \eb that varies slowly over the nuclear plane, {\em e.g.}, macroscopic terms like Eq.~\eqref{eq:SEMF},  approximately cancels in the GK relations. 
However, microscopic terms that exactly satisfy the full set of GK relations studied in Ref.~\cite{GARVEY:1969xgo} are severely restricted, they must factorize as in Eq.~\eqref{eq:mic-fac}. 
We henceforth refer to this as \textit{Jaffe factorization}. 

Dissecting the leading PCs $\mathcal{E}_{b,i}$ to determine which symbolic terms to use in a macroscopic component, $\eb^{\rm mac}$,
we find that the same modified version of Eq.~\eqref{eq:SEMF} used by the WS4 model, except replacing the Wigner term by 
$\eb^W = \alpha_W |Z-N|/A$, works well. 
This Wigner term has been considered in 
Refs.~\cite{Kirson:2008yvv,Moller:2015fba}, but is not included in many other popular macroscopic formulas.
Next, we consider a simple model using the macroscopic terms obtained from our study of the leading AI model PCs, along with Jaffe factorized microscopic terms as in Eq.~\eqref{eq:mic-fac}. 
We find that this simple model works remarkably well, except near the double-magic nuclei such as 
${}^{208}_{82}$Pb. 
The fact that Jaffe factorization breaks near these double-magic nuclei is not surprising,  so we adopt dedicated terms for these regions~\cite{Supp}.
After including these double-magic terms, our simple model achieves an RMS accuracy of 0.37~MeV; see Fig.~\ref{fig:res-symbolic}. 
Therefore, most of the improvement of the AI model over the macroscopic terms is captured.\footnote{Note that terms accounting for nuclear shell structure and magic numbers factorize, hence are implicitly included via Eq.~\eqref{eq:mic-fac}.}



\begin{figure}[b]
    \centering
    \includegraphics[width=0.44\linewidth]{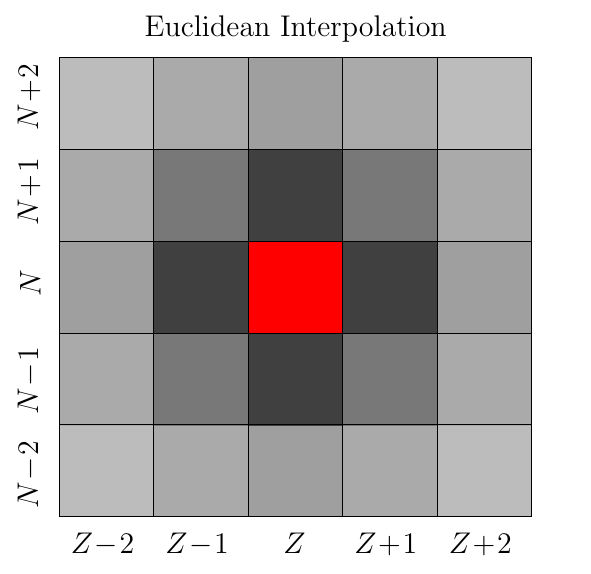} \hspace{-1.0em}
    \includegraphics[width=0.44\linewidth]{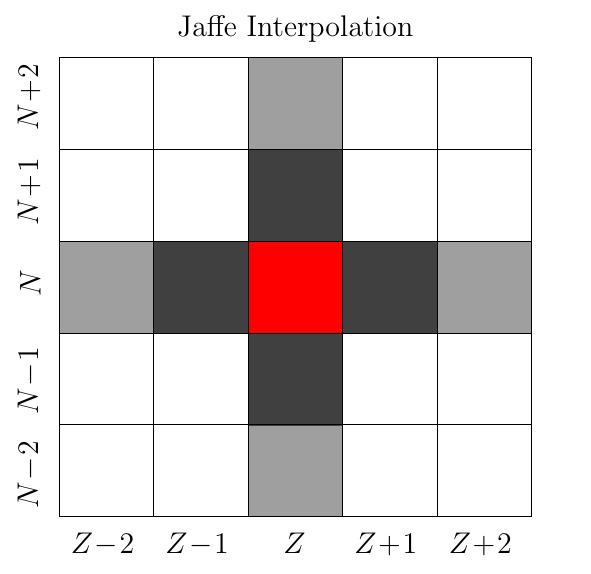} \hspace{-1.0em}
    \includegraphics[width=0.063\linewidth]{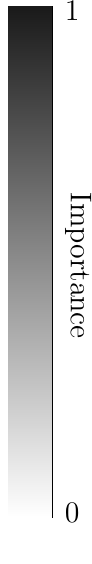}
    \vspace{-1.0em}
    \caption{Relative importance (shading)  of neighboring nuclei using (left) Euclidean (purely distance-based) and (right) Jaffe  interpolation (target nucleus at center, colored red.)}
    \label{fig:interp2}
\end{figure}

Jaffe factorization  in Eq.~\eqref{eq:mic-fac} is overly restrictive since the GK relations only hold {\em locally}, {\em i.e.}\ in small regions of the nuclear plane, whereas Eq.~\eqref{eq:mic-fac} applies global corrections for each $Z$ and $N$. 
We can relax Jaffe factorization and apply it locally using the observed \eb values of neighboring nuclei. 
Indeed, many models are based on learning local corrections to WS4 based on the distances to observed neighboring nuclei (see Fig.~\ref{fig:interp2} (left)), achieving excellent precision, {\em e.g.} an RMS of 0.18~MeV in Ref.~\cite{Wang:2014qqa}. 
However, local Jaffe factorization implies that distance in the nuclear plane is not the most important consideration.
Instead, neighbors with either the same $Z$ or $N$ are the most useful.
We find that corrections obtained from these neighbors, arranged as a $+$, not only outperform the remaining neighbors, arranged in an $\times$---but the $\times$ corrections provide no improvement~\cite{Supp}.\footnote{Jaffe factorization implies that the error incurred by using $\times$ neighbors is twice that of the $+$ neighbors~\cite{Supp}.} 
Even using next-to-nearest-neighbor $+$ nuclei dramatically outperforms nearest-neighbor $\times$ nuclei. 

\begin{figure}[t]
    \centering
    \includegraphics[width=1.0\linewidth]{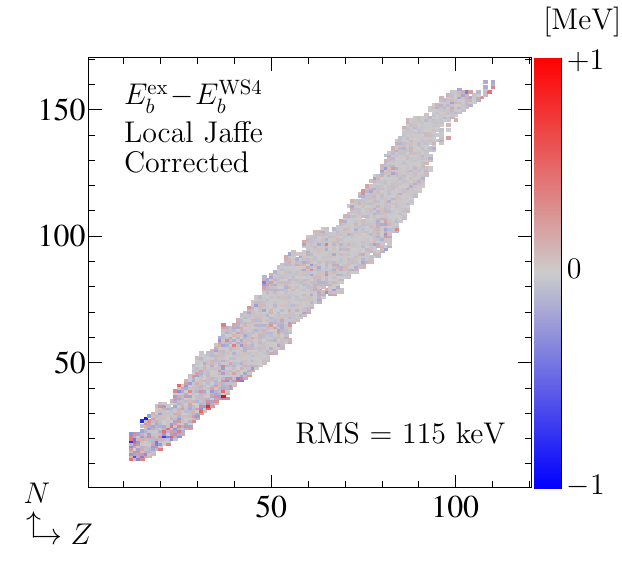}
    \vspace{-1.0em}
    \caption{Residuals of WS4 including local Jaffe corrections.}
    \label{fig:res-ws4-cor}
\end{figure}

Local Jaffe corrections using the $+$ nearest and next-to-nearest neighbors as in Fig.~\ref{fig:interp2} (right) reduces the RMS of our simple model, which recall has no physics microscopic terms,  to 0.18~MeV.
This is substantially better than WS4, the best physics model, and even consistent with
Ref.~\cite{Wang:2014qqa}.
Instead, applying these corrections to WS4 reduces its RMS to 0.16~MeV, which beats all but the most advanced AI models (such as Ref.~\cite{Kitouni:2023rct}) without using any AI or complicated interpolation.

Local Jaffe corrections can be improved by also correcting the neutronic (protonic) component for isotopic (isotonic) neighbors, which are used to obtain the protonic (neutronic) corrections in Fig.~\ref{fig:interp2} (right). 
This is done by considering all available nearest and next-to-nearest neighbors of nucleus $(Z,N)$ in a linear regression that determines local Jaffe corrections for each $Z \in [Z-2,Z+2]$ and $N \in [N-2,N+2]$~\cite{Supp}.\footnote{Local Jaffe corrections work well even when most neighboring nuclei are unobserved, which is a major advantage over using the GK relations directly; see Table~\ref{tab:localjafferms}.} 
Applying these corrections to  WS4 produces an RMS below 0.12~MeV (residuals shown in Fig.~\ref{fig:res-ws4-cor}). 
These are  state-of-the-art  results for \eb predictions---and they are based on explainable corrections to a physics model, not a black-box AI approach. 
%
Therefore, our main result is that AI has rediscovered a long forgotten nuclear property we refer to as Jaffe factorization, and when this is applied locally to physics-model calculations it provides state-of-the-art, interpretable predictions for nuclear masses.

In summary, AI models have shown promise making high-precision predictions of nuclear masses, some even surpassing the best physics models.
However, the reliability of AI predictions can be difficult to assess. 
We presented here the first AI model that not only achieves cutting-edge precision for $E_b$, but does so in an interpretable manner. 
We showed why the most important dimensions of its internal representation form a double helix, 
and that its predictions can be factorized and ordered hierarchically, with the most important terms corresponding to well-known symbolic models (such as the famous LD). 
Remarkably, the improvement of the AI model over symbolic ones can  be attributed to a nuclear property we refer to as localized Jaffe factorization.

The GK relations are the type of empirically derived patterns from data that an AI model is expected to learn, and the (local) Jaffe factorization solution is also something such models are good at discovering.
However, without the use of an architecture like ours---chosen to aid interpretability---it would have been difficult to realize what the AI had learned.
%
%
Given that we now know how optimal local corrections work in simple physics-based terms, it should be possible for nuclear theorists to better predict the highly unstable nuclei for which such corrections can be trusted.
Our work demonstrates how human-AI collaboration can unlock new scientific understanding and accelerate discovery.

\textit{Acknowledgments}---We are grateful to Bob Jaffe for informing us about his undergrad work, pointing us to the factorization presented in Ref.~\cite{GARVEY:1969xgo}, and providing comments on our work.  
We also thank Jesse Thaler, Antoine Belley, and Jose Munoz for providing useful feedback. 
This work was supported by NSF grant
PHY-2019786 (The NSF AI Institute for Artificial Intelligence and Fundamental Interactions, http://iaifi.org/). 
S.T. was supported by the Office of High Energy Physics of the US Department of Energy (DOE) under Grant No.~DE-SC0012567, and by the DOE QuantISED program through the theory consortium “Intersections of QIS and Theoretical Particle Physics” at Fermilab (FNAL 20-17).
S.T. is additionally supported by the Swiss National Science Foundation project number PZ00P2\_223581, and acknowledges CERN TH Department for hospitality while this research was being carried out.

\bibliography{main}

\input{supp}

\end{document}

%% file: supp.tex
\clearpage
\newpage

\onecolumngrid
\setcounter{equation}{0}
\setcounter{figure}{0}
\setcounter{table}{0}
\setcounter{section}{0}
\setcounter{page}{1}
\makeatletter
\renewcommand{\theequation}{S\arabic{equation}}
\renewcommand{\thefigure}{S\arabic{figure}}
\renewcommand{\thetable}{S\arabic{table}}
\pagestyle{plain}

\begin{center}
\textbf{\large The DNA of nuclear models: How AI predicts nuclear masses} \\
\vspace{0.05in}
{ \it \large Supplemental Material}\\
\vspace{0.05in}
{Kate A. Richardson,  Sokratis Trifinopoulos, and Mike Williams}
\end{center}

\section{Data Set \& Definition of Performance Metric}

We use data from the latest Atomic Mass Evaluation (AME2020) data set~\cite{Wang:2021xhn}. 
When quoting RMS errors, we focus on nuclei with $Z,N \geq 12$, since small nuclei are each distinct enough that it is difficult to infer their properties directly from other nuclei, {\em i.e.}\ without physics inputs.
We also remove from the data sample poorly measured nuclei---those with a precision worse than 100~keV or those determined with mass surface analysis--- since comparing to measured values with uncertainties larger than our target precision introduces unnecessary noise. 
This leaves 2325 nuclei with $12 \leq Z \leq 118$ and $12 \leq N \leq 178$. 

The RMS values we quote are unbiased, each nucleus is masked (not considered) when determining the model parameters used to predict its \eb value, except for simple models like LD which have only a small number of parameters where potential bias is inherently minimal (we also do not refit the WS4 model parameters).
This is especially important when comparing the performance of AI models. The AI model used in this study achieves an RMS on its training data of less than 10~keV, which only means that the model has sufficient capacity to fit the data. 
In Refs.~\cite{Kitouni:2023rct,Kitouni:2024ulw}, we quoted an RMS of 130~keV on all of the same nuclei used in this study. 
That is an unbiased RMS, which required independently training different versions of the AI model such that each \eb value is only considered an unbiased prediction when coming from a model trained without access to that nucleus. 
Importantly for a multi-task model, when removing $\eb(Z,N)$ this requires also removing from the training data the other properties of nucleus $(Z,N)$. In addition, all the neighboring separation energies must be removed, since $\eb(Z,N)$ can be exactly solved for using the right combination of neighboring separation energies. 

Unfortunately, not all AI models quote RMS values in the manner outlined here.
AI models that include the training data in the sample used to determine the RMS are (potentially highly) biased.
AI models that only consider a single training-testing split and quote an unbiased RMS using only the testing data  are only basing their RMS on a small sample of nuclei. 
This is troublesome since some nuclei are much more difficult to predict than others, and these are unlikely to be included in the RMS sample.
Multi-task AI models that are not careful to co-classify all inputs from the same nuclei in their training-testing split, {\em e.g.}\  labeling $\eb(Z,N)$ testing but the radius of $(Z,N)$ training, and instead classifying nuclei from each observable independently at random, are also biased. 
Consequently, this makes precise comparison of the performance of various AI predictions of \eb challenging. 
We hope that all future works will consider the definition of unbiased RMS described here.

\section{Details on Physics Models}

\subsection{Liquid Drop Model}

\begin{figure}[t]
    \centering
    \includegraphics[width=0.32\linewidth]{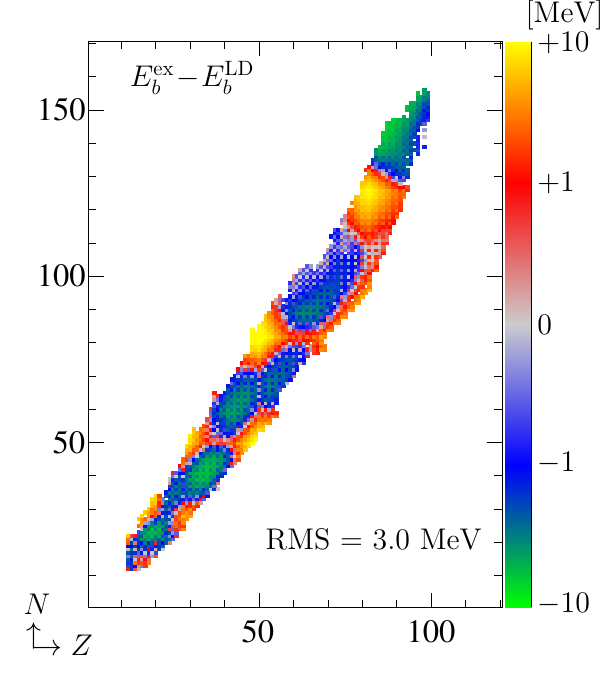}
    \hspace{0.5in}
    \includegraphics[width=0.32\linewidth]{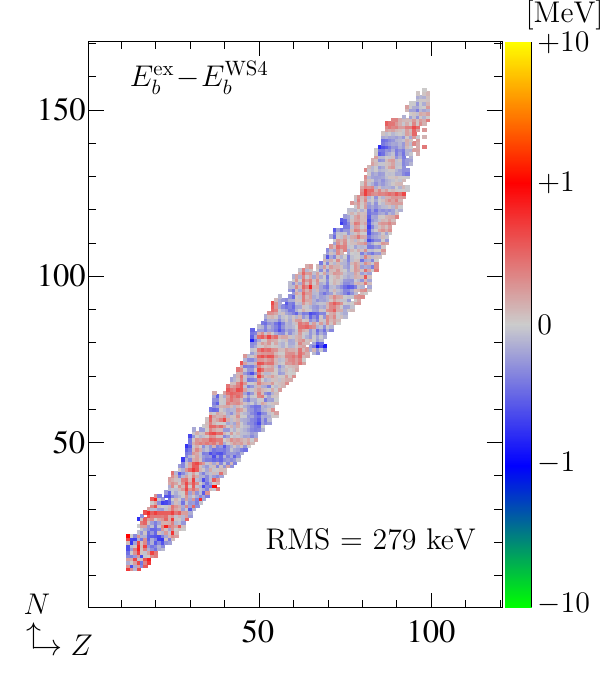}
    \caption{Residuals from the (left) LD and (right) WS4 models. Note that the color scale is linear on $[-1,1]$ and logarithmic otherwise. The lack of nuclear shell corrections is evident in the LD residuals.}
    \label{fig:res-LD-WS4}
\end{figure}

The famous liquid drop (LD) model, proposed almost a century ago~~\cite{Weizsacker:1935bkz,Bethe:1936zz}, treats the nucleus as a highly dense incompressible fluid. 
The nuclear binding energy, \eb, within this simple model is given by 
\begin{align}
\label{eq:SEMF-Supp}
\eb^{\rm LD} = \alpha_v A - \alpha_s A^{2/3} - \alpha_c\frac{Z(Z-1)}{A^{1/3}}  -\alpha_a \frac{(N-Z)^2}{A} + \alpha_p \frac{\delta(Z,N)}{A^{1/2}}\,,
\end{align}
where $A = Z+N$ is the total number of nucleons (protons and neutrons) in the nucleus. 
%
%
%
%
The terms in order are referred to as the volume, surface, Coulomb, asymmetry, and pairing terms.
For concreteness, when considering only the LD model, we use
\begin{align}
\label{eq:SEMFpairing}
\delta(Z,N) = \begin{cases}
    1 & $if $N$ and $Z$ even$\\
    -1 & $if $N$ and $Z$ odd$\\
    0 & $otherwise$
\end{cases}\,.
\end{align}
Despite its simplicity, after fitting the $\alpha_i$ coefficients to data, the LD predictions achieve a precision that ranges from $\mathcal{O}(\%)$ for moderate nuclei to $\mathcal{O}(0.1\%)$ for large nuclei. 
Figure~\ref{fig:res-LD-WS4} shows the residuals of the LD model across the nuclear plane.
The RMS for the LD model is roughly 3.0~MeV. 

\subsection{Weizs\"acker-Skyrme (WS4)}

To the best of our knowledge, version 4 of the Weizs\"acker-Skyrme (WS4) model~\cite{Wang:2014qqa} provides the best precision for a non-AI mode.
The WS4 model employs a microscopic-macroscopic approach.
The microscopic calculations use an axially deformed Woods–Saxon potential; see Ref.~\cite{Wang:2014qqa} for details.
The macroscopic model used by WS4 applies some corrections to the LD model. 
The volume and surface terms are unchanged, however the Coulomb term becomes 
\begin{align}
\label{eq:WS4Coulomb}
\eb^C = \alpha_C \frac{Z^2}{A^{1/3}} \left(1-\frac{0.76}{Z^{2/3}}\right)\,,
\end{align}
where only $\alpha_C$ is determined from a fit to the data. 
The asymmetry term becomes
\begin{align}
\label{eq:WS4sym}
\eb^{a} = \alpha_{a}\left(1-\frac{\kappa}{A^{1/3}}-\xi \frac{2-|I|}{2+|I|A}\right)I^2A\left(1+\kappa_{a}((I-I_0)^2-I^4)A^{1/3}\right)\,,
\end{align}
where $\alpha_{a}$, $\kappa$, $\xi$, and $\kappa_{a}$ are all parameters of the fit, $I = (N-Z)/A$, and $I_0 = 0.4A/(A+200)$. 
The version of the pairing term used is
\begin{align}
\label{eq:WS4pair}
\eb^{\rm pair} = \alpha_{\rm pair}A^{-1/3}\delta_{np}\,,
\end{align}
where
\begin{align}
\delta_{np} = \begin{cases}
    \frac{17(2-|I|-I^2)}{16} & $if $N$ and $Z$ even$\\
    |I| - I^2 & $if $N$ and $Z$ odd$\\
    1-|I| & $if $N$ even, $Z$ odd, and $N>Z\\
    1-|I| & $if $N$ odd, $Z$ even, and $N<Z\\
    1 & $otherwise$
\end{cases}\,,
\end{align}
and $\alpha_{\rm pair}$ is a parameter to be fit.
The final macroscopic term in the WS4 model is the Wigner term, which has the form
\begin{align}
\label{eq:WS4wigner}
\eb^W = \alpha_W(e^{|I|} - e^{-|\eta|})\,,
\end{align}
where $\alpha_W$ is found from fitting the data, $\eta = (\Delta Z \Delta N)/Z_m$, and $\Delta Z$ and $\Delta N$ are the number of nuclei between the nucleus of interest and the line defined by $N = 1.37 Z + 13.5$ along the Z and N directions, respectively. The quantity $Z_m$ is defined as the hypotenuse of the triangle formed by $\Delta Z$ and $\Delta N$.
Putting this all together, the predictions of the WS4 model are
\begin{align}
\label{eq:WS4}
\eb^{\rm WS4} = \alpha_v A - \alpha_s A^{2/3} + \eb^C + \eb^{a} + \eb^{\rm pair} + \eb^W + \eb^{\rm {mic,WS4}}\,,
\end{align}
where the final term denotes the WS4 microscopic corrections. 
Therefore, the macroscopic part is a symbolic expression with 9 parameters, while the microscopic part is the output of a simplified many-body quantum-mechanical calculation using a phenomenological potential with 10 parameters determined by fitting the data~\cite{Wang:2014qqa}. 
Figure~\ref{fig:res-LD-WS4} shows the residuals of the WS4 model across the nuclear plane.
The RMS for the WS4 model is 0.28~MeV.

\section{Details on the Architecture}

To facilitate extracting what knowledge the AI  has learned, we study a simplified version of the model in Refs.~\cite{Kitouni:2023rct,Kitouni:2024ulw} that is only trained to predict \eb, and thus, only has access to \eb measurements. 
This AI architecture takes as its only inputs for each nucleus the values of $Z$ and $N$, which are promoted to learnable 1024-dimensional vectors that are fed into a simple neural network (NN), $F_{\rm{nn}}$, with learnable parameters $\vec{\theta}$:
\begin{align}
Z \to \vec{Z}, N \to \vec{N} \Rightarrow \eb = F_{\rm{nn}}(\vec{Z},\vec{N},\vec{\theta})\,.
\end{align}
The first step in the architecture involves concatenating together $\vec{Z}$ and $\vec{N}$ into a single vector with $n = n_Z+n_N$ components.
Schematically, the NN is depicted in Fig. \ref{fig:architecture}.
%
The ResBlocks 
are essentially just $n \to n$ multi-layer-perceptrons plus residual (or skip) connections:
\begin{align}
 {\rm ResBlock}(\vec{x}) = \vec{x} + \left[ \vec{x} \to {\rm Linear}(n,n) \to {\rm ReLU} \to {\rm Linear}(n,n) \to {\rm ReLU} \right]\,.
\end{align}
The objective during training is to minimize the following loss function:
\begin{align} \label{eq:loss_app}
\mathcal{L} = \sum_i \left(E_{b,i}^{\rm ex} -E_{b,i}^{\rm ai}\right)^2 + \lambda \left[ \sum_j Z_j^2 + \sum_k N_k^2 + \sum_{\ell} \theta_{\ell}^2  \right]\,, 
\end{align}
where $E_{b,i}^{\rm ex}$ and $E_{b,i}^{\rm ai}$ are the experimental and AI-predicted \eb values for each nucleus in the training data, and $\lambda$ is a hyperparameter that controls the balance between prediction quality and model complexity. 
This hyperparameter is introduced via the Adaptive Moment Estimation (Adam) algorithm, and in our case the variant AdamW~\cite{2017arXiv171105101L}, which is used during training with learning rate $10^{-4}$ and weight decay $10^{-1}$ for over 150,000 epochs.

Finally, we note that while we use the basis $(Z,N)$, it might seem more natural, given Eq.~\eqref{eq:SEMF}, to instead use $(A,Z-N)$.
Note that the form of the regularization term in Eq.~\eqref{eq:loss_app} is such that both bases give the same penalty term, which means that this alternative basis is just a rotation from our chosen one.
For Refs.~\cite{Kitouni:2023rct,Kitouni:2024ulw} we chose the $(Z,N)$ basis due to our expectation that the nuclear magic numbers would be special in some way in the embedded structures. 
Indeed, that is seen in Refs.~\cite{Kitouni:2023rct,Kitouni:2024ulw}, but not seen here, due to those prior works also training on separation energies where the shell structure is more important. 


\tikzset{
  box/.style = {draw, minimum width=2.2cm, minimum height=1.2cm, align=center, thick},
  smallbox/.style = {draw, minimum width=1.2cm, minimum height=1cm, align=center, thick},
  arrow/.style = {ultra thick, ->, >=Stealth, draw=teal},
  resblock/.style = {smallbox, fill=gray!20},
  embbox/.style = {smallbox, fill=cyan!20},
  outputbox/.style = {smallbox, fill=orange!30}
}

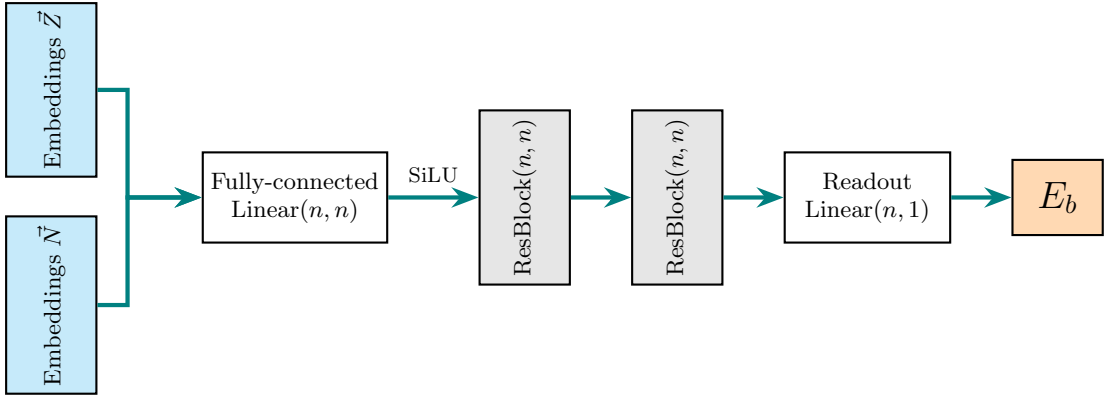
\begin{figure}[t]
\centering
\begin{tikzpicture}[node distance=0.5cm and 0.8cm]

\node[embbox] (zemb) {\rotatebox{90}{Embeddings $\vec{Z}$}};
\node[embbox, below=of zemb] (nemb) {\rotatebox{90}{Embeddings $\vec{N}$}};

\node[box, right=2cm of $(zemb)!0.5!(nemb)$] (fc) {Fully-connected \\ Linear$(n,n)$};

\node[resblock, right=1.2cm of fc] (res1) {\rotatebox{90}{ResBlock$(n,n)$}};
\node[resblock, right=of res1] (res2) {\rotatebox{90}{ResBlock$(n,n)$}};

\node[box, right=of res2] (readout) {Readout \\ Linear$(n,1)$};
\node[outputbox, right=of readout] (eb) {\Large{$E_b$}};

\draw[arrow] (zemb.east) -- ++(0.4,0) |- (fc.west);
\draw[arrow] (nemb.east) -- ++(0.4,0) |- (fc.west);

\draw[arrow] (fc) -- node[midway, above, yshift=0.2em, font=\footnotesize, black] {SiLU} (res1);
\draw[arrow] (res1) -- (res2);
\draw[arrow] (res2) -- (readout);
\draw[arrow] (readout) -- (eb);

\end{tikzpicture}
\caption{The neural network model architecture. Inputs are embeddings of $Z$ and $N$, which are
concatenated and then processed through a fully-connected layer and two residual blocks, followed by a readout predicting the binding energy $E_b$.}
\label{fig:architecture}
\end{figure}

\section{Details on the Formation of a Double Helix}

\subsection{Setting the Scale in the Representation Space}

The dominant term in Eq.~\eqref{eq:SEMF} is the volume term, thus $\eb \approx \alpha_v A$. 
The AI model must be able to build $A$ from \pczv and \pcnv. 
This drives the most important PC components to be 
\begin{align}
\label{eq:volume-approx}
\pcz_1 \approx \beta Z, \pcn_1 \approx \beta N \to A \approx (\pcz_1 + \pcn_1)/\beta\,,
\end{align}
where $\beta$ is a learned scale factor. 
Addition of \pczv and \pcnv components is trivial to implement in the downstream NN. 
Therefore, to produce the volume term the NN just needs to store the factor $\alpha_v/\beta$ in its $\vec{\theta}$ parameters. 
The regularization term in Eq.~\eqref{eq:loss_app} provides a \textit{central force} that attracts all degrees of freedom in the AI model inwards. 
Naively, this should drive $\beta \to 0$; however, this necessarily results in a contribution from the $\vec{\theta}$ regularization of $\alpha_v/\beta \to \infty$. 
Consider the simplified case 
\begin{align}
\eb = \alpha_v A, \pcz_1 = \beta Z, \pcn_1 = \beta N, F_{\rm{nn}}(\vec{Z},\vec{N},\vec{\theta}) = \frac{\alpha_v}{\beta}(\pcz_1 + \pcn_1) = \alpha_v A\,.
\end{align}
The scale $\alpha_v/\beta$ must be encoded as a sum of products of $\theta$ parameters, which we will denote generically as $F_{\Sigma\Pi}(\vec{\theta})$, with the constraint that $F_{\Sigma\Pi}(\vec{\theta}) = \alpha_v/\beta$. 
The form of $F_{\Sigma\Pi}(\vec{\theta})$ depends on the architecture, but the constraint will always be holonomic. In the limit of the best-fit point, {\em i.e.} where the first term in Eq. \eqref{eq:loss_app} vanishes, the loss at the minimum satisfies
\begin{align}
\vec{\nabla} \mathcal{L} = \Lambda \vec{\nabla} \left(F_{\Sigma\Pi}(\vec{\theta}) - \frac{\alpha_v}{\beta}\right),
\end{align}
where $\Lambda$ denotes a Lagrange multiplier and the gradient is with respect to all learnable parameters of the AI model. 
The optimal loss in this case satisfies 
\begin{align}
\mathcal{L} = \lambda \left[ \beta^2 (s_Z + s_N) + \sum_\ell \theta_{\ell}^2 \right], \qquad s_Z = \sum_{Z} Z^2, \qquad s_N = \sum_{N} N^2\,, 
\end{align}
where $s_{Z,N}$ are geometric sums of the (unique) values of $Z$ and $N$ found in the training data set, hence $s_{Z,N}$ are constants that do not vary during training. 
The value of $\beta$ that minimizes the loss is thus
\begin{align}
\beta = \left[ \frac{\Lambda\alpha_v}{2\lambda(s_Z + s_N)} \right]^{1/3}\,,
\end{align}
where the value of $\Lambda$ depends on $F_{\Sigma\Pi}(\vec{\theta})$, and in general $\Lambda \propto \beta^{\gamma}$ for some power $\gamma$. 
(For our architecture, the optimal solution is either $\gamma = 0$ or $1/2$ depending on the values of $\alpha_v$, $\lambda$, and $(\sigma_Z + \sigma_N)$.)
We can see that the constraint of obtaining the natural scale $\alpha_v$ results in determining a (non-zero) scale for the PCs, {\em i.e.}, $\beta \neq 0$. 
We confirm this by taking a frozen model, {\em i.e.} a model where the neurons of all downstream layers are kept constant, and rescale the first PC components, which as expected gives
\begin{align}
\pcz_1 \to (1\pm\delta)\pcz_1, \pcn_1 \to (1\pm\delta)\pcn_1 \Rightarrow \eb \to (1\pm\delta)\alpha_v A\,.
\end{align}
Therefore, the compressive regularization pressure is counter-balanced by the goodness-of-fit term in $\mathcal{L}$, which effectively produces a \textit{repulsive force} when the embedding vectors get too close along the first PC axis (similar to the Van der Waals force in DNA). 
The scale $\beta$ is set by the natural scale $\alpha_v$, along with details about the architecture and training data.

%
%
%
%
%
%

\subsection{Forming Helices}

The same reasoning that prevents the first PC from collapsing under the regularization pressure also applies to the second and third PC dimensions shown in Fig.~\ref{fig:DNA}; we will show this later in this section.
However, this does not explain the origins of the helix structure.
Naively, there are no cyclic or oscillatory terms in Eq.~\eqref{eq:SEMF}, making it surprising that the second and third most important PCs would form an oscillatory structure.

To understand why an oscillatory structure is optimal here, we next consider the second most important term in Eq.~\eqref{eq:SEMF}, the asymmetry term, which contains a factor of $(Z-N)^2$.
This term is invariant under transformations of the form $Z \to Z+\gamma, N \to N+\gamma$, which correspond to translations along the $\hat{\pcz}_1$ and $\hat{\pcn}_1$ axes (which are nearly aligned) in Fig.~\ref{fig:DNA}. 
What matters when computing  $(Z-N)^2$ is only the relative difference in $Z$ and $N$; thus, the same solution can be repeated all along the first PC axis, which is desirable due to the regularization pressure.
In principle our AI model could learn to store $(Z-N)^2$ for each $(Z,N)$ pair, but the regularization pressure forces it to instead find a more efficient solution, and the helix is the most efficient option. 

\begin{figure}[t]
    \centering
    \includegraphics[width=0.5\linewidth]{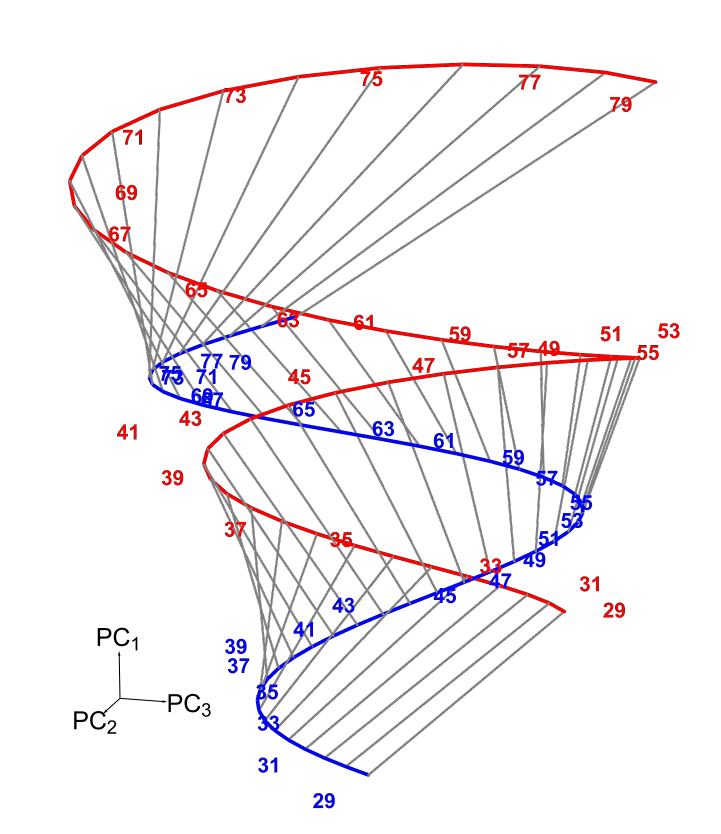}
    \caption{The three most important PCs of the internal representations learned by our AI model for (red) $Z$ and (blue) $N$
    when trained on simulated isospin-symmetric data.
    The links connect the values of $Z$ and $N$ found in the most stable nucleus of each isotopic chain, which is $Z=N$ for this simulated data.  
    The curves show fits of helices to the $Z$ and $N$ representations. Note that only the odd points are labeled to avoid clutter.}
    \label{fig:DNA-toy}
\end{figure}

As in the previous subsection, it is illustrative to consider a simplified scenario.
We generate isospin-symmetric training data using 
\begin{align}
\eb = \alpha_v A - \alpha_a \frac{(Z-N)^2}{A}\,,
\end{align}
and enforce that the observed nuclei are symmetric in the sense that for any generated nucleus $(Z,N)$ the mirror nucleus $(N,Z)$ is also generated. 
Figure~\ref{fig:DNA-toy} shows that a nearly isospin-symmetric double helix is formed in this scenario.
Given the dominance of the volume term, we again find the relationship from Eq.~\eqref{eq:volume-approx}. 
The focus here will be on PC dimensions 2 and 3, namely the second and third most important PCs. 

\begin{figure}[t]
    \centering
    \includegraphics[width=1\linewidth]{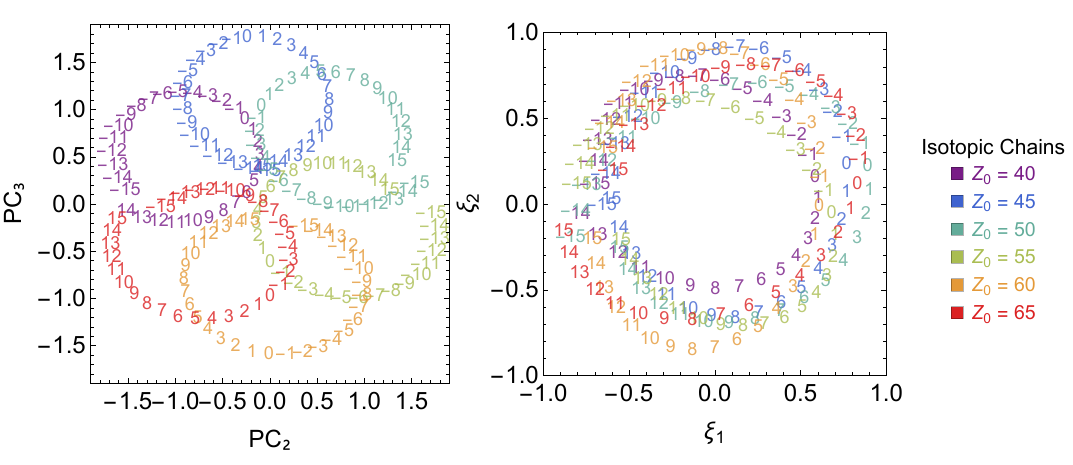}
    \caption{Left: Embedding vector components $(\pcn_2 - \pcz_2,\pcn_3 - \pcz_3)$ linking  each point $Z_0$ on the $Z$ helix with the points $[N_0-15,N_0+15]$ with $N_0=Z_0$ on the $N$ helix. 
    The numbers for each labeled point denote $N-N_0 = N - Z_0$. Right: Transformation of the points in the left panel according to Eq. \eqref{eq:clock-vec}.}
    \label{fig:clock-ex}
\end{figure}

In the limit of a perfect double helix in the first 3 components of both  \pczv and \pcnv, we first note that the number of isotopes (equivalently isotones), $n_I$, is related to the angular frequency of points along the helix by
\begin{align}
\omega \approx 2\pi/n_I\,.
\end{align}
Next, consider the following vector in the $\rm{PC}_2$-$\rm{PC}_3$ plane:
\begin{align}
\label{eq:clock-vec}
\vec{\xi} = \begin{pmatrix}
\pcz_2 \pcn_2 - \pcz_3 \pcn_3 \\
\pcz_2 \pcn_3 + \pcz_3 \pcn_2 
\end{pmatrix}\,.
\end{align}
Figure~\ref{fig:clock-ex} shows that, in this isospin-symmetric scenario, for any isotope (considering all $\vec{\xi}$ for a fixed $Z_0$)
\begin{align}
\vec{\xi}(Z_0,N) \propto \begin{pmatrix}
\cos{[(Z_0-N)\omega]} \\
\sin{[(Z_0-N)\omega]} 
\end{pmatrix}\,,
\end{align}
where a similar mirror relationship exists for isotones. (This relationship is not exact due to effects such as the $\pcz_1$ and $\pcn_1$ axes not being truly parallel, and some leakage of information into the lesser PC components.)
Evidently, the simple construction Eq.~\eqref{eq:clock-vec} encodes the value of $(Z-N)$ needed to build the asymmetry term. 

A transformer-based model could easily construct $\vec{\xi}$ using the dot product operator provided by the attention mechanism. (Notice that Eq.~\eqref{eq:clock-vec} involves products of components of the embedding vectors.)
However, our model does not use attention.
It is straightforward though to build up $\vec{\xi}$ instead using only sums of embedding components and nonlinearities in the NN. 
See Appendix~A of Ref.~\cite{zhong2023clock} for a derivation showing how to obtain $\vec{\xi}$ from only sums and nonlinearities.
Of note, whereas  $\vec{\xi}$ as defined in Eq.~\eqref{eq:clock-vec} is proportional to the radius of the helices squared, $\rho^2$, since the components in Eq.~\eqref{eq:clock-vec} involve products of $\pcz_{2,3}$ and $\pcn_{2,3}$ components, for the case where attention is not used this proportionality changes to $|\vec{\xi}| \propto \rho$. 
As in the previous subsection, the regularization pressure naively drives $\rho \to 0$, but again the need to predict the scale of the asymmetry term, driven by the constant $\alpha_a$, leads to an optimal solution where $\alpha_a$ is encoded as $\rho$ multiplied by a sum of products of $\theta$ parameters. 
We confirmed this by taking a frozen model and rescaling $\rho$, which gives
\begin{align}
\rho \to \beta \rho \Rightarrow \eb \to \alpha_v A - \beta \alpha_a  \frac{(Z-N)^2}{A}\,,
\end{align}
as expected for this architecture.

Finally, we note that the electromagnatic interactions of protons breaks the symmetry between $Z$ and $N$, which is why the double helix in Fig.~\ref{fig:DNA} is itself not symmetric. 
For example, the most stable nuclei in nature do not have $Z=N$; the Coulomb term in Eq.~\eqref{eq:SEMF} requires many more neutrons than protons in large stable nuclei.
The nuclear shell structure is also an important factor in determining which nucleus is the most stable for each isotopic chain.
These result in the bond analogs in Fig.~\ref{fig:DNA} not being parallel like they are in Fig.~\ref{fig:DNA-toy} (especially close to magic nuclei). 

\section{Physics Origins of Jaffe Factorization}

The Garvey-Kelson (GK) relations are patterns found in nuclear data, where combinations of \eb from a set of nuclei  approximately cancel almost anywhere in the nuclear plane~\cite{GARVEY:1969xgo}. 
For example, if $Z > N$,
\begin{align}
\eb(Z\!+\!2,N\!-\!2) -  \eb(Z\!+\!2,N\!-\!1)  - \eb(Z\!+\!1,N\!-\!2) 
  + \eb(Z,N\!-\!1)  + \eb(Z\!+\!1,N) - \eb(Z,N)  \approx 0 \, ,
\end{align}
whereas if $ Z < N$,
\begin{align}
\label{eq:gk1}
\eb(Z\!-\!2,N\!+\!2) -  \eb(Z\!-\!1,N\!+\!2)  - \eb(Z\!-\!2,N\!+\!1) 
  + \eb(Z\!-\!1,N)  + \eb(Z,N\!+\!1) - \eb(Z,N)  \approx 0 \, .
\end{align}
Other GK relations also approximately hold, such as
\begin{align}
\label{eq:gk2}
\eb(Z,N\!+\!2) -  \eb(Z\!-\!2,N)  + \eb(Z\!-\!2,N\!+\!1) 
  - \eb(Z\!-\!1,N\!+\!2)  + \eb(Z\!-\!1,N) - \eb(Z,N\!+\!1)  \approx 0 \, .
\end{align}
The GK relations arise because single-nucleon energy levels do not vary much in a small region of the nuclear plane. 
Therefore, in any such small region around the point $(Z,N)$ we can write
\begin{align}
\label{eq:Eb-fact}
\eb(Z+\delta Z,N + \delta N) \approx  \sum_i^{Z+\delta Z} E^p_i(Z,N) + \sum_j^{N+\delta N} E^n_j(Z,N) \, ,
\end{align}
where $E^{p,n}_{i,j}(Z,N)$ denote the single-(proton,neutron) energy levels for nuclei near the point $(Z,N)$. 
Notice that each energy level appears equally with a plus and minus sign in the GK relations; therefore, plugging Eq.~\eqref{eq:Eb-fact} into any GK relation gives zero under the assumption that the single-nucleon energy levels are constants within the region considered. 
In the limit that the GK relation holds exactly, we thus have
\begin{align}
\label{eq:Jaffe-fact-local}
\eb(Z+\delta Z,N + \delta N)  \Rightarrow F_Z(Z+\delta Z) + F_N(N + \delta N) \, ,
\end{align}
which is Jaffe factorization. 
Any contribution to \eb that varies slowly over the nuclear plane, {\em e.g.}, macroscopic terms like Eq.~\eqref{eq:SEMF},  approximately cancels in the GK relations, but not exactly, hence this factorization approximation only holds in small local regions of the nuclear plane.  
Finally, note that in Ref.~\cite{GARVEY:1969xgo} two non-factorizable terms also appear, including a constant multiplied by $ZN$. 
Following the derivation presented here, that constant only need not vary over the small region around $(Z,N)$, {\em e.g.}, it could be chosen to be $-2\alpha_a/A$ and the $ZN$ term could thus be absorbed into the asymmetry term from Eq.~\ref{eq:SEMF}. 
The other non-factorized term presented in Ref.~\cite{GARVEY:1969xgo} similarly can be absorbed into the pairing term. 
Clearly we can also factor out any constants from $F_Z+F_N$, and absorb them into the macroscopic model as well as other slowly varying contributions to \eb. 
This is why we only focus on the factorized expression.

\begin{figure}[b]
    \centering
    \includegraphics[width=0.35\linewidth]{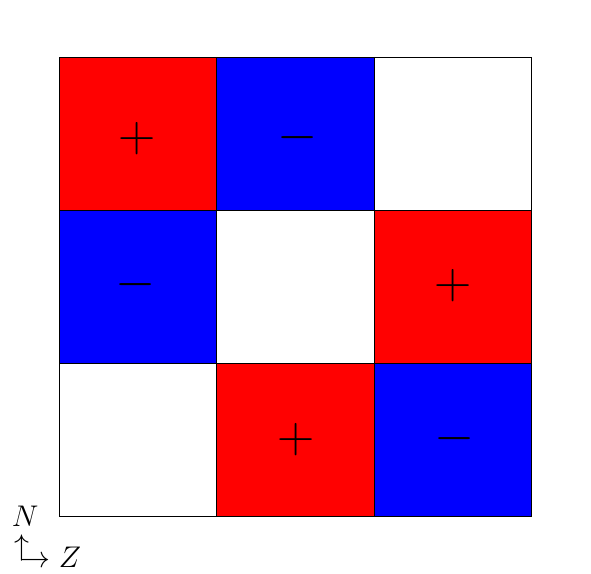} 
    \hspace{1.0em}
    \includegraphics[width=0.35\linewidth]{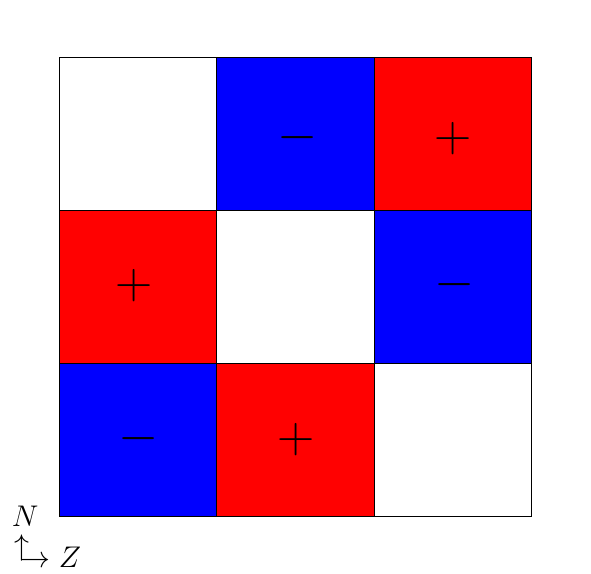} 
    \caption{Examples of GK relations in the $(Z,N)$ plane for Eqs.~\eqref{eq:gk1} and \eqref{eq:gk2}. The relative sign (addition or subtraction) of the contribution of each \eb value is labeled on the panel (also colored red for addition, blue for subtraction).  }
    \label{fig:GK}
\end{figure}

\section{Penultimate-Layer Principle Components \& Our Symbolic Model}

Figure~\ref{fig:penultimate-pcs} shows the 6 most important PCs of the penultimate NN layer, where the sum of all such PCs is $\eb^{\rm ai}$. 
The first PC is dominated by the volume term.
In fact, the first two PCs when summed are well approximated by Eq.~\eqref{eq:SEMF} with the exception of what are typically called shell terms (more on these below).
The third PC visually has an approximate linear dependence on $Z-N$, and we find it cannot be fit well without the Wigner term described in the main text.
The lesser PCs all exhibit Jaffe factorization to a high degree, such that the RMS from fitting the sum of all lesser PCs to Eq.~\eqref{eq:Eb-fact} is around 0.1~MeV. 
Note, however, that the fifth PC also has a strong overlap with the pairing term (which itself approximately obeys Jaffe factorization).

We build the macroscopic portion of our symbolic model by fitting the sum of the first 3 PCs, which are approximately smooth functions.
This base model largely follows the one used by WS4, but with a few modifications. 
In addition to the terms defined above, our model includes corrections near the double-magic nuclei ${}^{56}_{28}$Ni, ${}^{78}_{28}$Ni, ${}^{100}_{50}$Sn, ${}^{132}_{50}$Sn, and ${}^{208}_{82}$Pb of the form 
\begin{align}
\label{eq:doublemagic}
\eb^{\rm DM} = \alpha_{\rm DM}e^{-\beta_{\rm DM}\Delta}\,,
\end{align}
where $\Delta$ is the Euclidean distance to the double-magic nucleus. 
We also include a microscopic portion that consists of the Jaffe relations as described above by Eq.~\eqref{eq:mic-fac}, since even the first 3 PCs do have discrete Jaffe-like terms built into them (mostly describing nuclear shell structure). 
To be explicit, the full model is 
\begin{align}
\label{eq:symbolic_model}
\eb^{\rm sym} = \alpha_v A - \alpha_s A^{2/3} + \eb^C + \eb^{a} + \eb^{\rm pair} + \alpha_W\frac{|Z-N|}{A} + \eb^{\rm DM} + \eb^{\rm mic,Jaffe}\, .
\end{align}
The volume and surface terms are unchanged from Eq.~\eqref{eq:SEMF}. The Coulomb, asymmetry, and pairing terms are from Eqs.~\eqref{eq:WS4Coulomb}--\eqref{eq:WS4pair}, and the BW2 Wigner term~\cite{Kirson:2008yvv} is as described in the main text. These are combined with 249 constants, one for each $N$ and $Z$ value for the global Jaffe corrections. 
As discussed in the main text, the model is improved by replacing the global Jaffe terms with local ones; these are described in the following sections. 

Some care must be taken when fitting this model since the global Jaffe corrections have some degeneracy with the macroscopic terms and each other. For example, the volume term can be factorized according to Eq.~\eqref{eq:mic-fac}, and thus can be absorbed into the $\eb^{\rm mic,Jaffe}$ term, setting $\alpha_V =0$. While this results in equivalent predictions, it reduces interpretability and removes physical meaning from the model. Similarly, the Jaffe corrections can be degenerate with each other, {\em e.g.} by adding a constant term to $F_Z$ and subtracting it from $F_N$. To deal with these two forms of degeneracy, we include Gaussian penalties around the parameters in the first 6 terms of Eq.~\ref{eq:symbolic_model} and L1 regularization on the microscopic parameters in the loss for the fit. The full loss to be minimized is thus
\begin{align}
\label{eq:model_loss}
\mathcal{L} \!=\! \sum_i \left(E_{b,i}^{\rm ex} \!-\! E_{b,i}^{\rm sym}\right)^2\! + \lambda_G\sum_i \left(\frac{\alpha^{\rm fit}_i - \bar{\alpha}_i}{\bar{\alpha}_i}\right)^2 + \lambda_R\sum_i|\theta_i^J|\,,
\end{align}
where $E_{b,i}^{\rm ex}$ and $E_{b,i}^{\rm sym}$ are the experimental and predicted $\eb$ values for each nucleus, $\lambda_G$ and $\lambda_R$ are hyperparameters which control the strictness of the Gaussian and L1 constraints, $\alpha_i^{\rm fit}$ and $\bar{\alpha}_i$ are the parameters fit in the full model or only macroscopic version, and $\theta_i^J$ is the set of microscopic corrections.

\begin{figure}[t]
    \centering
    \includegraphics[width=0.32\linewidth]{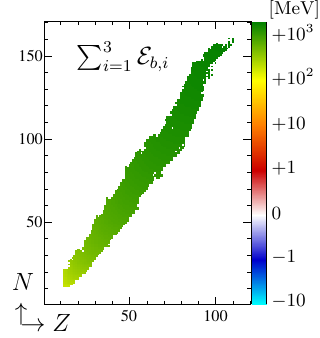} 
    \includegraphics[width=0.32\linewidth]{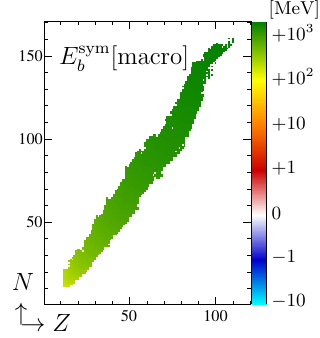} 
    \includegraphics[width=0.32\linewidth]{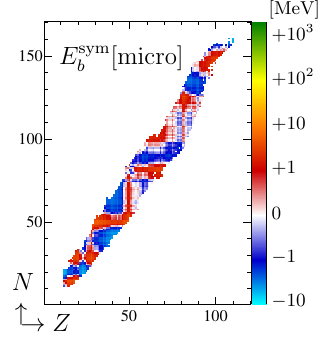} 
    \caption{Left: Sum of the 3 most important PCs in the penultimate NN layer. Middle: Macroscopic model obtained by fitting the data shown in the left panel. Right: Microscopic model obtained from the same fit using global Jaffe terms. The predominant features observed in these microscopic terms are the shell structure near nuclear magic numbers.}
    \label{fig:PCfits}
\end{figure}

Figure~\ref{fig:PCfits} shows the results of fitting the first 3 PCs to Eq.~\eqref{eq:symbolic_model}.
The first 3 PCs are described well, and the shell structure contained in them is clearly visible in the microscopic part of the function.
Figure~\ref{fig:res-symbolic} shows the residuals from fitting Eq.~\eqref{eq:symbolic_model} to the data. 
The RMS is 0.37~MeV using the global Jaffe terms. 
Given the simplicity of this model, its accuracy is impressive. 
Using the local Jaffe corrections reduces the RMS substantially as well, as described below and in the main text. 
Local Jaffe corrections are discussed in detail in the next section. 

\section{Jaffe Factorization \& Local Corrections}

A common approach to using AI to improve the precision of \eb predictions is to start from the WS4 model and learn how to interpolate and extrapolate corrections for each nucleus based on the residuals of its neighboring nuclei.
A simplified non-AI version of this approach is to use nearest-neighbor and possibly next-to-nearest-neighbor nuclei to derive the correction as follows:
\begin{align}
\eb = \eb^{\rm WS4} + \sum_i \frac{\left(E_{b,i}^{\rm ex} - E_{b,i}^{\rm WS4}\right)}{\sqrt{(Z-Z_i)^2 + (N - N_i)^2}} \cdot \left[ \sum_i \left((Z-Z_i)^2 + (N - N_i)^2  \right)^{-1/2} \right]^{-1}\,,
\end{align}
where the sum runs over a set of neighboring nuclei.
Kernels other than the inverse distance give nearly identical results, hence only the inverse-distance kernel is shown here. 
Applying such corrections gives the following results. 
\vspace{0.5em}

\noindent \textbf{No corrections}: The WS4 model without any neighbor-based corrections has an RMS of 279~keV. \vspace{0.5em}

\noindent \textbf{All nearest neighbors}: Using all  nuclei with $|Z-Z_i| \leq 1$, $|N - N_i| \leq 1$, and $(Z,N) \neq (Z_i,N_i)$ to derive the local correction reduces the RMS to 207~keV. \vspace{0.5em}

\noindent \textbf{Only isoto(p,n)ic nearest neighbors}: Using only nuclei with either $Z = Z_i$ and $|N - N_i| = 1$ or $N = N_i$ and $|Z - Z_i| = 1$ to derive the local correction, which is half the nuclei from the previous case, does even better, reducing the RMS to 175~keV. \vspace{0.5em}

\noindent \textbf{Only non-isoto(p,n)ic nearest neighbors}: Conversely, using nuclei with $|Z-Z_i| = 1$ and $|N - N_i| = 1$  to derive the local correction does much worse than the previous case, despite using the same number of nuclei, yielding an RMS of 277~keV, consistent with the uncorrected WS4 model. \vspace{0.5em}

\noindent \textbf{Only isoto(p,n)ic next-to-nearest neighbors}: Using  nuclei with either $Z = Z_i$ and $|N - N_i| = 2$ or $N = N_i$ and $|Z - Z_i| = 2$ to derive the local correction gives an RMS of 186~keV. Even though these nuclei are all farther away than the non-isoto(p,n)ic nearest neighbors, they provide a much better correction. \vspace{0.5em}

\noindent \textbf{Only isoto(p,n)ic (next-to-)nearest neighbors}: Using nuclei with either $Z = Z_i$ and $1 \leq |N - N_i| \leq 2$ or $N = N_i$ and $1 \leq |Z - Z_i| \leq  2$ to derive the local correction gives an RMS of 160~keV. This value is an improvement over the much more complicated approach using radial basis vectors in Ref.~\cite{Wang:2017fhd}. \vspace{0.5em}

\noindent \textbf{Only non-isoto(p,n)ic (next-to-)nearest neighbors}: Conversely, nuclei with with $1 \leq |Z-Z_i| \leq 2$ and $1 \leq |N - N_i| \leq 2$  to derive the local correction does much worse than the previous case, despite using twice as many nuclei, producing an RMS of 242~keV. \vspace{0.5em}



These results demonstrate the efficacy of Jaffe factorization~\cite{GARVEY:1969xgo} applied to local corrections, as illustrated in Fig.~\ref{fig:interp2}. 
Naively, the closer in the nuclear plane a given nucleus is to the target nucleus, one would expect it gives more information about how to correct the WS4 prediction.
However, this is not the case. 
The isoto(p,n)ic neighbors provide extremely valuable information to use in deriving corrections to WS4, whereas the non-isoto(p,n)ic neighbors do not (directly, see following section).
The problem is not purely geometric in the nuclear plane due to the quantum-mechanical nature of nuclei (see discussion in the \textit{Physics Origins of Jaffe Factorization} section). 

\begin{figure}[t]
    \centering
    \includegraphics[width=0.32\linewidth]{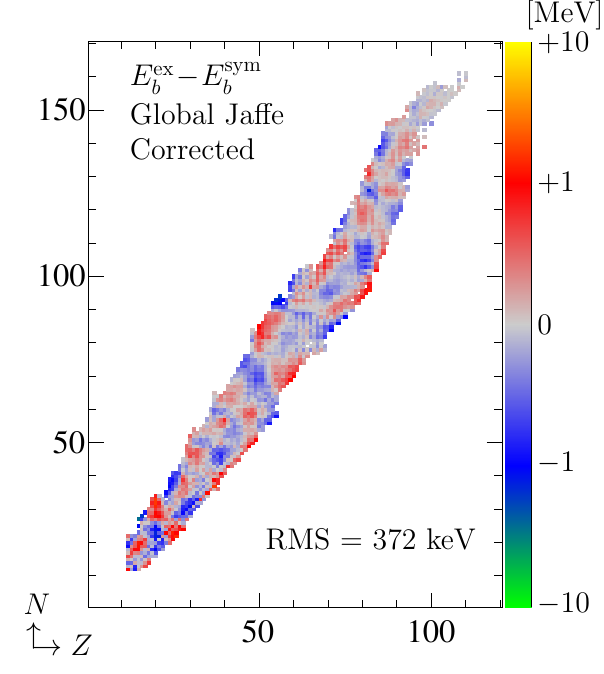}
    \includegraphics[width=0.32\linewidth]{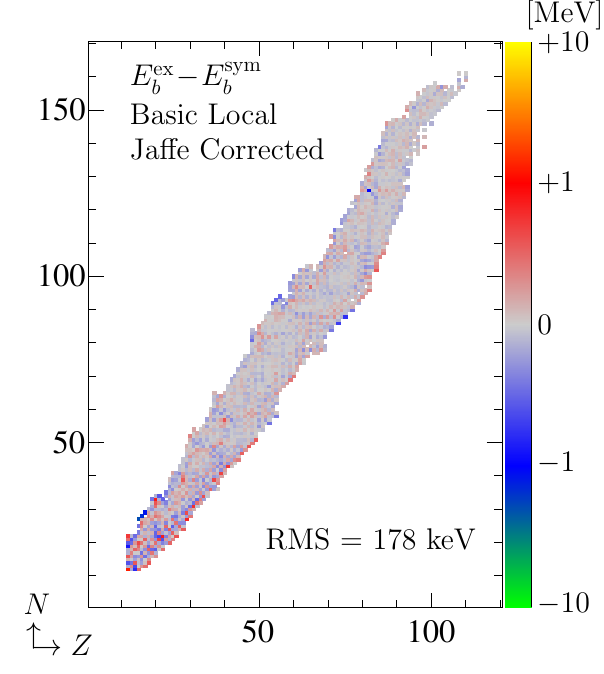}
    \includegraphics[width=0.32\linewidth]{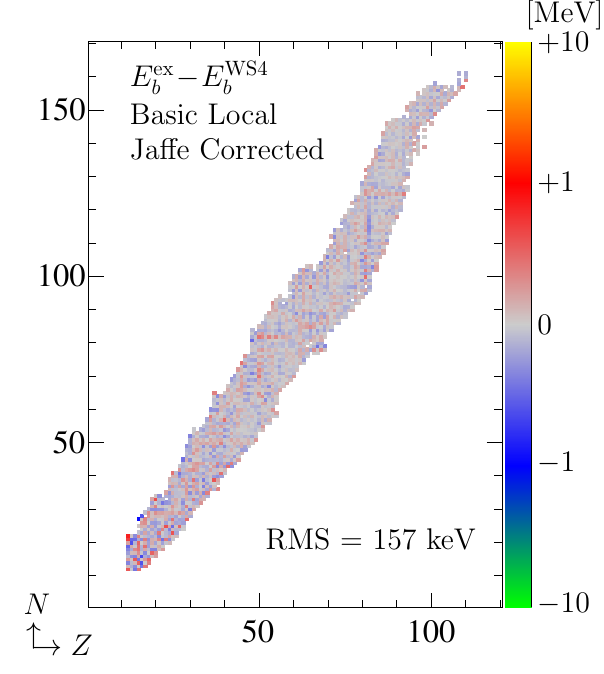}
    \caption{Residuals from our symbolic model with (left) global and (middle) basic local Jaffe corrections, along with (right) the WS4 model with basic local Jaffe corrections. Here, basic Jaffe refers to corrections using the scheme as shown in Fig.~\ref{fig:interp2} (right). Note that the color scale is linear on $[-1,1]$ and logarithmic otherwise. }
    \label{fig:res-symbolic}
\end{figure}

We previously showed that Jaffe factorization arises as a local approximation based on the fact that the single-nucleon energy levels do not vary much within a small region of the nuclear plane.
Equation~\eqref{eq:Jaffe-fact-local} is a direct consequence of this approximation, 
and using it to derive the value of $\eb(Z,N)$ from only its isoto(p,n)ic nearest neighbors gives
\begin{align}
\eb(Z,N) &\approx \frac{1}{4}\left[ F_Z(Z) + F_N(N-1) + F_Z(Z) + F_N(N+1) + F_Z(Z-1) + F_N(N) + F_Z(Z+1) + F_N(N) \right] \nonumber \\
& = \frac{1}{2} \left[ F_Z(Z) + F_N(N) \right]  + \frac{1}{2} \left[ \frac{F_Z(Z-1) + F_Z(Z+1)}{2} + \frac{F_N(N-1) + F_N(N+1) }{2} \right] \\
& = F_Z(Z) + F_N(N) + \delta_Z + \delta_N \nonumber \, . 
\end{align}
The approximation made to get to the third line is exact if $F_Z$ and $F_N$ are linear functions, with the errors induced by nonlinearities in $F_{Z,N}$ denoted $\delta_{Z,N}$. 
Therefore, this local Jaffe interpolation is valid up to non-linear corrections required when changing $Z$ or $N$ by one unit. 
Using only the non-isoto(p,n)ic neighbors instead gives
\begin{align}
\eb(Z,N) &\approx \frac{1}{4}\left[ F_Z(Z\!-\!1) \!+\! F_N(N\!-\!1) \!+\! F_Z(Z\!-\!1)\! +\! F_N(N\!+\!1) \!+\! F_Z(Z\!+\!1)\! +\! F_N(N\!-\!1) \!+\! F_Z(Z\!+\!1) \!+\! F_N(N\!+\!1) \right] \nonumber \\
& = \frac{F_Z(Z-1) + F_Z(Z+1)}{2} + \frac{F_N(N-1) + F_N(N+1) }{2} \\
& = F_Z(Z) + F_N(N) + 2\left( \delta_Z + \delta_N \right)\nonumber \, . 
\end{align}
Notice that this approximation does not contain $F_Z(Z) + F_N(N)$ directly like the previous one does---and is twice as sensitive to non-linear errors. 
The isoto(p,n)ic neighbors share the same filled proton or neutron energy levels as the target nucleus, with only one vacant or extra filled neutron or proton energy level; whereas the non-isoto(p,n)ic neighbors always have a difference in occupancy of energy levels for both the protons and neutrons. 
This explains why Jaffe corrections vastly outperform distance-based methods in deriving local corrections for nuclear masses. 

The physics of local Jaffe corrections can be illustrated by redoing the derivations of the previous paragraph starting instead from Eq.~\eqref{eq:Eb-fact}.
For the isoto(p,n)ic nearest neighbors this gives
\begin{align}
\eb(Z,N) &\approx \frac{1}{4}\left[ 2\sum_{i=1}^{Z} E_i^p + \sum_{j=1}^{N-1} E^n_j + \sum_{j=1}^{N+1} E^n_j + \sum_{i=1}^{Z-1} E^p_i + 2\sum_{j=1}^{N} E^n_j + \sum_{i=1}^{Z+1} E^p_i \right] \nonumber \\
& = \sum_{i=1}^{Z} E_i^p + \sum_{j=1}^{N} E^n_j + \frac{1}{4}\left[(E_{Z+1}^p - E^p_Z) + (E_{N+1}^n - E^n_N)  \right] \, .
\end{align}
We see that the errors due to nonlinearities, $\delta_{Z,N}$, arise from differences in the single-nucleon energy levels. 
Repeating this calculation for the non-isoto(p,n)ic neighbors gives the same result but with $1/4 \to 1/2$ in the last term, as expected since as already shown the $\times$ interpolation is twice as sensitive to nonlinearities as the $+$ interpolation.

Finally, we find a small improvement by not using  odd-odd $Z=N$ neighbors when deriving the local Jaffe corrections.
In addition, we do not attempt to correct odd-odd $Z=N$ nuclei. 
These improvements, {\em e.g.},reduce the RMS from 160 to 157~keV for the (next-to-)nearest isoto(p,n)ic neighbors.  
These small modifications are included in Fig.~\ref{fig:res-symbolic}, though the impact on the RMS is small.

\section{Improving on Basic Local Jaffe Corrections}

Consider the up to 24 nuclei made up of all observed nearest and next-to-nearest neighbors of a target nucleus $(Z,N)$.
The isotopic neighbors used in the local Jaffe corrections from the previous section induce some error due to the implicit assumption that the  discrepancy  between WS4 and data is only due to the protonic component of \eb. 
Similarly, the isotonic neighbors introduce an error due to the assumption that the discrepancy is only due to the neutronic component. 
We can instead take each \eb value to locally be
\begin{align}
\eb(Z,N) \approx \eb^{\rm WS4}(Z,N) + \alpha_Z + \alpha_N \,, 
\end{align}
where $\alpha_Z$ and $\alpha_N$ are corrections to the protonic and neutronic WS4 energy levels near the nucleus $(Z,N)$. 
Previously, we obtained $\alpha_Z$ and $\alpha_N$ using the distance-weighted averages of the isotopic and isotonic neighbors (local Jaffe corrections as in Fig.~\ref{fig:interp2}).
We can improve on this by instead solving for these corrections by minimizing 
\begin{align}
\chi^2 = \sum_{Z',N'} \mathbf{1}[(Z',N') \neq (Z,N)]  \left( \eb^{\rm ex} - \eb^{\rm WS4} - \alpha_{Z'} - \alpha_{N'} \right)^2 \, , \, |Z-Z'| \leq 2\, ,\, |N-N'| \leq 2\, , 
\end{align}
where $\mathbf{1}$ is an indicator function that excludes the target nucleus from the sum (since its \eb value cannot be used in its own prediction), and of course, also excludes all unobserved nuclei. 
This approach allows us to simultaneously correct the neutronic (protonic) component for isotopic (isotonic) neighbors while using them to obtain the protonic (neutronic) correction. 
Note that as in the basic local Jaffe corrections described previously, we find a small improvement by excluding neighbors on the opposite side of the $Z=N$ line. 

Applying this procedure---except for $Z=N$ odd-odd nuclei, and nuclei with less than 7 total and 3 isoto(p,n)ic measured neighbors (of the 24 possible), where we instead apply the basic Jaffe procedure from the previous section---to the WS4 predictions reduces the RMS down to 115~keV.
The residuals are shown in Fig.~\ref{fig:res-ws4-cor}. 
We are unaware of any unbiased RMS values that are this low, including from advanced AI models. (Note that some AI works include training-data nuclei in their RMS numbers, which can be highly biased.)

Table~\ref{tab:localjafferms} shows how our procedure compares with using the GK relations directly to correct WS4.
For nuclei that have nearly all nearest and next-to-nearest neighbors observed, the local Jaffe procedure is similar to GK using a mean of all 12 GK-type predictions for each nucleus~\cite{Barea:2008zz}.
The real advantage of our procedure over using the GK relations directly is that local Jaffe corrections also work well when most of the neighboring nuclei are not measured, which is the most interesting region.
%
%
As noted in the Letter, scientifically what is desired is not higher-precision predictions of nuclei surrounded by well-measured neighbors, but trustworthy ways of extrapolating corrections away from the stable region.
The local Jaffe approach provides an interpretable method for deriving such corrections. 

\begin{table}[t]
\centering
\begin{tabular}{c|c|c|c}
Number (Next-to-)Nearest Neighbors & WS4 RMS & GK Corrected RMS & Local Jaffe Corrected RMS \\ \hline
20--24 & 268 keV & 93 keV & 94 keV \\
11--19 & 300 keV & 161 keV & 160 keV \\
1--10 & 391 keV & 293 keV & 207 keV 
\end{tabular}
\caption{Performance of the WS4 model without (middle) and with (right) local Jaffe corrections for nuclei categorized by how many of their (next-to-)nearest neighbors are included in the data sample. If all of the (next-to-)nearest neighbors are observed, the value is 24. As expected the performance of the WS4 model, along with both GK and our local Jaffe corrections degrades when moving towards the edges of the observed nuclear plane; however, even when most of the neighbors are not observed, the local Jaffe procedure still drastically improves the precision of the nuclear mass predictions. In addition, our Local Jaffe corrections outperform the GK-based corrections for nuclei with few observed neighbors. }
\label{tab:localjafferms}
\end{table}